\UseRawInputEncoding
\documentclass[aps,prd,floatfix,superscriptaddress,preprintnumbers,nofootinbib,twocolumn,longbibliography,letterpaper,natbib,nofootinbib]{revtex4-1}
\pdfoutput=1
\usepackage{amsmath, amssymb, amsfonts, graphicx, mathrsfs, verbatim, float, slashed, bbm, color, xcolor, xspace, enumerate, url, bbding, multirow, outlines,upgreek}
\usepackage[english]{babel}
\usepackage[colorlinks=true,linkcolor=black,citecolor=galacticcenterbubblegum]{hyperref}
\definecolor{galacticcenterbubblegum}{rgb}{0.8,0, 0.8}

\usepackage{graphicx,bm}

\newcommand{\beq}{\begin{equation}}
\newcommand{\bea}{\begin{eqnarray}}
\newcommand{\eeq}{\end{equation}}
\newcommand{\eea}{\end{eqnarray}}
\newcommand{\bal}{\begin{align}}
\newcommand{\eal}{\end{align}}

\usepackage{tikz}
\usetikzlibrary{decorations.pathmorphing,decorations.markings}
\tikzset{
photon/.style={decorate, decoration={snake,amplitude=4pt, segment length=7pt}, draw=black},
particle/.style={draw=black, postaction={decorate}, decoration={markings,mark=at position .5 with {\arrow[draw=black]{>}}}},
antiparticle/.style={draw=black, postaction={decorate}, decoration={markings,mark=at position .5 with {\arrow[draw=black]{<}}}},
gluon/.style={decorate, draw=black, decoration={coil,amplitude=3pt, segment length=4pt}},
higgs/.style={draw=black,dashed,thick },
arrow/.style={draw=black, very thick, postaction={decorate}, decoration={markings,mark=at position 1 with {\arrow[draw=black]{>}}}}
}

\usepackage{latexsym}
\usepackage{subcaption}

\newcommand{\MeV}{~\mathrm{MeV}}
\newcommand{\GeV}{~\mathrm{GeV}}

\usepackage{outlines}
\usepackage{enumitem}
\setenumerate[1]{label=\Roman*.}
\setenumerate[2]{label=\Alph*.}
\setenumerate[3]{label=\roman*.}
\setenumerate[4]{label=\alph*.}

\newcommand{\Lc}{\mathcal{L}}

\newcommand{\G}{\text{G}}

\newcommand{\T}{T_\star}
\newcommand{\R}{\text{R}}
\newcommand{\M}{\text{M}}
\newcommand{\Tx}{T_{\text{cf}}}
\newcommand{\TEW}{T_{\text{EW}}}

\newcommand{\Mpl}{M_{\text{Pl}}}
\newcommand{\Gphi}{\Gamma_\Phi}

\definecolor{darklightsabergreen}{rgb}{0.0, .49, 0.06}

\begin{document}
\title{Magnetogenesis From Baryon Asymmetry During an Early Matter Dominated Era } 
\preprint{MITP-21-059}
\author{Fatemeh Elahi}
\email{felahi@uni-mainz.de}
\affiliation{Kavli Institute for the Physics and Mathematics of the Universe (WPI), UTIAS, The University of Tokyo, Kashiwa, Chiba 277-8583, Japan
}
\affiliation{PRISMA$^+$ Cluster of Excellence \& Mainz Institute for Theoretical Physics, \\
Johannes Gutenberg-Universität Mainz, 55099  Mainz, Germany} 
\author{Hadi Mehrabpour}
\email{hadi.mehrabpour.hm@gmail.com}
\affiliation{Frankfurt Institute for Advanced Studies, Giersch Science Center, D-60438 Frankfurt am Main, Germany}
\affiliation{Kavli Institute for the Physics and Mathematics of the Universe (WPI), UTIAS, The University of Tokyo, Kashiwa, Chiba 277-8583, Japan
}

\vspace*{0.5cm}
\begin{abstract}
\vspace*{0.5cm}

{In this paper, we study the simultaneous evolution of baryon asymmetry and hypermagnetic field amplitude assuming an early matter domination. We contrast our results to the conventional case where radiation domination during early universe is assumed. We show that the baryon asymmetry and the hypermagntic field amplitude can change by orders of magnitude if we assume a non-standard history of cosmology. That is because the Hubble rate determines which processes are efficient. We find that a change in Hubble rate can have a significant impact on when the weak sphalerons become active. As a result of a change in the evolution of baryonic asymmetry, alters the evolution of hypermagnetic field amplitude. It is known that if the hypermagnetic field amplitude is  large enough, it can save the baryon asymmetry from diminishing. We show that whether a small seed of hypermagnetic field amplitude can be amplified to a large enough value will strongly depend on the history of cosmology. 

 }
\end{abstract}
\maketitle
\section{ Introduction}
\label{sec:Intro}

According to our knowledge of the Big Bang Nucleosynthesis (BBN)~\cite{Cooke:2013cba} and the Cosmic Microwave Background (CMB)~\cite{Ade:2015xua}, we expect the universe to be in the stage of radiation domination (RD) around the temperature of $(10\MeV - \text{eV})$. On the other hand, we know that the energy densities of radiation, matter, and dark energy evolve as $ a^{-4}$, $a^{-3}$, and $a^0$ respectively, with $a$ being the scale factor. Therefore, the most natural and simplistic assumption about the early universe is that it was in the stage of RD after reheating up until BBN and CMB. In this assumption, if we have any matter particle with a sizable mass, it decays instantly once the temperature falls below its mass. However, it is possible to have a matter particle $\Phi$ with a relatively long lifetime before it decays. In this case, the assumption of instantaneous decay of $\Phi$ is no longer valid and for some interval in temperature, the energy density of $\Phi$ dominates the energy density of the universe~\cite{Vilenkin:1982wt}. The most important consequence of early matter domination (EMD) is that the Hubble rate is faster:
\begin{align}
& H_{\M} \equiv \sqrt{\frac{8 \pi (\rho_{\R} + \rho_\Phi)}{3 \Mpl ^2}},
\end{align}
where $\rho_{\R}$ is the energy density of radiation, $\rho_\Phi$ is the energy density of $\Phi$, and $\Mpl = 1.2 \times 10^{19} \GeV$ is the Planck mass. If $ \rho_\Phi \gg \rho_{\R}$, we are in the era of matter domination.  The phenomenological importance of early matter domination (EMD) scenario was first realized long ago in the context of supersymmetry ~\cite{Moroi:1999zb,Vilenkin:1982wt}. However, it has gained a lot of attention recently, especially to revive some of the well-motivated dark sector proposals which had not survived the experimental scrutiny ~\cite{Chanda:2019xyl,Chanda:2021tzi,Randall:2015xza,Kane:2015jia,Patwardhan:2015kga,Berlin:2016gtr,Berlin:2016vnh} (for more details see~\cite{Allahverdi:2020bys} and references therein).  If one of these models is correct and we are indeed living in a universe with such cosmological history, it is important to analyze the effects of EMD on other phenomena. Specifically, since the Hubble rate determines which processes are efficient in early universe, a change in Hubble rate can have important phenomenological outcomes.  One place that this feature can have significant effects is in the abundance of baryon asymmetry, which is the focus of this paper. 

The value of matter-antimatter asymmetry has been measured independently by BBN~\cite{Cooke:2013cba} and the Planck~\cite{Ade:2015xua}
 to \beq
\eta_B \equiv \frac{n_B - n_{\bar B}}{s} \simeq 8.5 \times 10^{-11},
\eeq
with $s = 2\pi^2 g_\star T^3/45$  being the entropy density, and $g_\star $ is  the relativistic degrees of freedom.

A study that wants to explain baryogenesis must respect the famous Sakharov conditions: 1) Baryon number violation, 2) C and CP violation, and 3) out of thermal equilibrium process ~\cite{Sakharov:1967dj}.  Even though in the Standard Model (SM) we are aware of a CP violation in weak interactions, the amount of the CP violation is not enough to explain the baryon asymmetry. Among the mentioned requirements, only the first one is satisfied in the SM. Baryon (and Lepton) symmetries are two of the known anomalous accidental symmetries of the SM. The baryon number in the SM is violated through the interactions of the weak sphalerons, which are highly active before the Electroweak Phase Transition (EWPT), but have a suppressed interaction rate after the EWPT. Among the SM fermions, right-handed electrons are special, because they have the smallest Yukawa couplings. Therefore, if right-handed electrons acquire an asymmetry, they will keep it relatively longer, until their Yukawa interaction enters thermal equilibrium ($\sim 10^4 -10^5 \GeV$)\cite{Dvornikov:2011ey, Semikoz:2015wsa}. After which, the asymmetry from right-handed electrons will be transferred to left-handed electrons. Then, weak sphalerons will distribute the asymmetry between all left-handed particles and eventually wash out the asymmetry~\cite{Klinkhamer:1984di}.  Studies have shown that in the standard cosmology, if we start even with a large asymmetry in the right-handed electron ($\mu/T \sim 1$, with $\mu$ being the chemical potential), then the sphalerons still have enough time to eat up the asymmetry to a much smaller value than observed~\cite{Zadeh:2018usa}. Therefore, some extra handles are needed. One way is to employ non-standard cosmology~\cite{Chen:2019wnk}.

If we have matter domination in the early universe, the Hubble rate will increase. Therefore, the temperature at which the electron Yukawa interaction exceeds the Hubble rate is smaller than the standard cosmology case. That will give sphalerons less time to eat up the asymmetry. Notice that in this case, the matter domination should not be too long such that the electron Yukawa enters equilibrium after EWPT. In other words, we do need the sphalerons to distribute the asymmetry, but we need to give them \textit{less} time to eat up the asymmetry. 

Similar to weak sphalerons, the Abelian anomaly may also lead to fermion number violation before EWPT: $\partial_\mu j^\mu_F \sim \frac{g^{'2}}{4\pi^2} \vec E_Y \cdot \vec B_Y$, with $g'$ being the hypercharge and $\vec E_Y$ ($B_Y$) being the hyperelectric (hypermagnetic) field\cite{Giovannini:1997eg,Giovannini:1997gp}. Before EWPT hyperelectric and hypermagnetic field couple to chiral fermions, whereas after EWPT, the ordinary electromagnetic field's coupling to fermions is vector-like. Hence, this fermion number violating effect is important only before EWPT.  The intimate relationship between hypermagnetic field and baryon asymmetry has been discussed in the literature~\cite{Semikoz:2011, Dvornikov:2013, Kuzmin:1985mm, Zadeh:2018usa,Long:2014,Rubakov:1986am,Giovannini:1998,Giovannini:1998b,Joyce:1997, Khlebnikov:1988sr,Zadeh:2016nfk,Zadeh:2015oqf,Mottola:1990bz}. Abelian anomaly can lead to fermion number violation only if we have a non-zero seed of hypermagnetic field in the early universe. Interestingly, the observation of magnetic field with similar amplitudes and wave length in causally disconnected patches of the universe may also indicate that there was a primordial seed of hypermagnetic field in the early universe~\cite{Kronberg:1993vk,Semikoz:2011, Dvornikov:2013,Quashnock:1988vs, Kibble:1995,Sigl:1997,Vachaspati:1991nm,Enqvist:1993,Enqvist:1994,Olesen:1997,Baym:1996,Grasso:2001,Neronov:2010,Neronov:2009,Tavecchio:2011,Tavecchio:2010,Giovannini:1998b,Joyce:1997,Wolfe:2008,Kulsrud:2008,Harrison:1973zz}. Even though there are some uncertainties in the evolution of the hypermagnetic field amplitude, a rough estimate suggests that if the amplitude of the hypermagnetic field is on the order of $10^{20} \G$ by the time of EWPT, the amplitude of the observed magnetic fields can be justified~\cite{Dvornikov:2013,Fujita:2016,Giovannini:1998b,Giovannini:1998,Giovannini:2013,Long:2016,Joyce:1997}. In this paper, we will consider the effect of the hypermagnetic field on the evolution of baryon asymmetry in the cases of both standard cosmology and a universe with an EMD. The Boltzmann equations describing the evolution of baryon asymmetry and hypermagnetic field is coupled, and therefore, we do a numerical study of a few benchmarks. To gain some intuition about the solutions, we present (an approximate) analytical solutions to some of the evolution equations. We show that their behavior in standard cosmology (SC) and EMD can be vastly different and non-intuitive. We emphasize that we are oblivious to the origin of EMD. Since EMD arise naturally in many beyond the SM solutions, we merely focus on how the evolution of baryon asymmetry and hypermagnetic field amplitude will change in the era of early matter domination.

 The organization of the paper is as the following: In Sec~\ref{sec:EDM}, we explain the basics of early matter domination; Sec~\ref{sec:BG} is dedicated to the evolution of baryon asymmetry. Specifically, subsection~\ref{sec:prime} is the evolution of $\eta_B$ assuming a primordial asymmetry in the right-handed electron. The subsection~\ref{sec:primeBY} is with the same assumption plus a non-zero seed of hypermagnetic field amplitude. In subsection~\ref{sec:gradual}, we show the evolution of $\eta_B$ and $B_Y$ when the asymmetry in the right-handed electron is from a gradual decay of another particle. Finally, the concluding remarks are presented in Sec.~\ref{sec:con}.

\section{Early Matter Domination}
\label{sec:EDM}
According to the SC, we expect to be in the era of radiation domination after reheating and until CMB. However, it is theoretically motivated to assume an early matter domination (EMD) that affects the Hubble rate for a limited range of temperature~\cite{Moroi:2000,Vilenkin:1982wt,Coughlan:1983ci,Starobinsky:1994bd,Dine:1995uk,Chung:1998rq,Giudice:2000ex,Allahverdi:2020bys,Chanda:2019xyl,Chanda:2021tzi,Maldonado:2019qmp,Guo:2020grp,Chang:2021ose,Dalianis:2020gup,Cosme:2020nac,Bernal:2020ywq}. It is important to note that the dominating particle has to decay so that by the time of BBN we are in the radiation-dominated epoch again. There are several ways to get EMD. For simplicity and for the sake of working with an explicit model, let us introduce a weakly interacting scalar $\Phi$ with a decay width $\Gamma_\Phi$. Because $\Phi$ is weakly interacting, then it goes through coherent oscillation and thus redshifts as $a^{-3}$. Hence, it is conceivable that at $\T > m_\Phi$, the energy density of $\Phi$ dominates over radiation\footnote{That is assuming $v_\Phi \gg \T$ such that $ v_\Phi^2 m_\Phi^2 \simeq \rho_R$ at $\T$. This assumption is reasonable in the context of many motivated extensions of the Standard Model (e.g, Ref.~\cite{Co:2019wyp,Chen:2019wnk}). Furthermore, it is important that the production of $\Phi$ freezes out at $T> \T$, so that $\Phi$ stays in a \textit{coherent} oscillation.}. In other words, $\T$ is defined when $\rho_\Phi = \rho_\R$.  The Boltzmann Equations (BEs) describing the evolution of $\Phi$ and radiation are the following:
\begin{align}
&\dot \rho_\Phi + 3 H_\M \rho_\Phi = - \Gamma_\Phi \rho_\Phi,\nonumber\\
&\dot \rho_{\R} + 4 H_\M \rho_{\R} = \Gamma_\Phi \rho_\Phi,
\label{eq:Hubble}
\end{align}
with $H_\M =\sqrt{ \frac{8\pi}{3 \Mpl^2} (\rho_\Phi + \rho_\R)}$, being the Hubble rate, and $\Gamma_\Phi$ is the total decay width of $\Phi$. The decay of $\Phi$ is radiation, and therefore increases the energy density of radiation. Eventually, $\rho_\Phi$ becomes negligible compared with $\rho_R$ and we return to radiation domination. The approximate analytical solution to Eq.~\ref{eq:Hubble} in the regime $t_\star < t < \Gamma_\Phi^{-1}$ is the following~\cite{Rubakov:2017xzr}:
\begin{align}
\rho_\Phi^a &\simeq \frac{\Mpl^2}{6 \pi t^2} e^{- \Gamma_\Phi t} \nonumber\\
\rho_R^a& \simeq \frac{\ Mpl^2 t_\star^{2/3}}{6 \pi t^{8/3}} + \frac{\Gamma_\Phi \Mpl^2}{10 \pi t},
\label{eq:analyticalrhos}
\end{align}
where $t_\star$ is the time corresponding to $\T$.

This phenomenon leads to a dilution factor that has been discussed extensively in the literature. In this work, however, we want to highlight some of the other consequences that have been given less attention but lead to non-trivial effects on the abundance of baryonic and dark matter. 

\begin{figure}
\centering 
\includegraphics[width=0.45\textwidth, height=0.22\textheight]{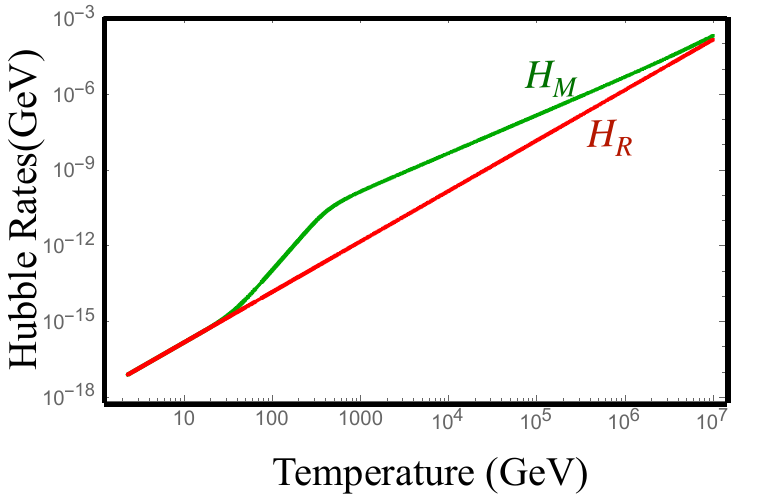}
\caption{The Hubble rates  $ H_\M$ (shown in green) and in the standard cosmology $ H_\R$ (red) are shown. As the plot demonstrates, the Hubble rate is increased until $\Phi$ decays into radiation, after which the two Hubble rates match ($\rho_\Phi \ll \rho_\R$). This plot is with the assumption that $ \T =  10^7 \GeV$  and $\Gamma_\Phi = 5 \times 10^{-15} \GeV$. }
\label{fig:Hubbles}
\end{figure} 

\section{Baryogenesis Abundance}
\label{sec:BG}

Explaining the matter-antimatter asymmetry has been the focus of many studies(e.g,~\cite{Affleck:1984fy,Fukugita:1986hr,Luty:1992un,Cohen:1993nk,Flanz:1994yx,Linde:2005ht,Dine:1995uk,Riotto:1999yt,Grasso:2000wj,Morrissey:2012}). If the fermionic sector gains an asymmetry before the Electroweak Phase Transition (EWPT), weak sphalerons become one of the key elements in the dynamics of the asymmetry. Weak sphalerons act on left-handed chiral fermions and rapidly eat the asymmetry preserving $B-L$ with an interaction per unit volume proportional to $\alpha_w^5 T^4$, where $\alpha_w$ is the coupling constant of the weak interaction. After EWPT, however, this rate is exponentially suppressed, and weak sphalerons become inactive\cite{Arnold:1996dy,DOnofrio:2012yxq}.

Requiring weak sphalerons to preserve $B-L$ limits its domain of interaction. More specifically, conserving $B-L$ and requiring the plasma to be neutral\footnote{The plasma needs to be neutral of  $SU(3) \times SU(2)\times U(1)_Y$ charges.}, leads to some equilibrium conditions among the chemical potential of fermions (see Appendix~\ref{app:EC} for more details). Therefore, the chemical potentials of SM fermions are related to each other. If we have an asymmetry in one of the fermions, we have an asymmetry in the rest of the fermions as well. In particular, we can express the chemical potential of baryons in terms of the chemical potential of the right-handed electron: $\mu_B = \frac{198}{481} \mu_{e_R}$\cite{Cline:1993bd,Chen:2019wnk}. Therefore, if we only study the BE of the right-handed electron, we can find the asymmetry of the baryons.  

Right-handed electron because of its limited interactions plays a crucial role in keeping the asymmetries. It only participates in hypermagnetic\footnote{The hypermagnetic interaction will also play an important role in the dynamics of the right-handed asymmetry, which we will discuss later.} and Yukawa interaction. That means that the strong and weak sphalerons cannot act on right-handed electrons. Furthermore, the Yukawa coupling of the right-handed electron is the smallest\footnote{That is if we ignore the neutrinos for now.}; thus, the chirality flip of the electrons freezes into equilibrium the last. In standard cosmology, the temperature at which the rate of chirality flip of the electron ($\Gamma_{LR}$) matches the Hubble rate $\Gamma_{LR} = H_\R$ is denoted by $\Tx$. The rate of the chirality flip of the electron has been estimated by Refs.~\cite{Campbell:1992,bodeker2019equilibration,Kamada:2016}: $\Gamma_{LR}  \simeq 10^{-2} y_e^2 T$. Hence, in the SC, $\Tx \simeq 10^4-10^5 \GeV$~\cite{Dvornikov:2011ey, Semikoz:2015wsa}. The small Yukawa coupling of the electron tames the weak sphalerons to wash out the asymmetry until $\Tx$. However, it is shown that the difference between $\Tx$ and $\TEW$, in SC, is long enough that the weak sphalerons can still wash out the asymmetry significantly~\cite{Campbell:1992,Cline:1993,Cline:1994}. In the following, we will discuss how changing the history of early universe cosmology can affect the phenomenology of baryon asymmetry. 

\subsection{Primordial Asymmetry }
\label{sec:prime}

In this subsection, we assume that the right handed electron has a large primordial asymmetry and we study the evolution of its asymmetry in SC and EMD. More specifically, we assume $\mu_{e_R}^{0}=  \T/2$, where the superscript $0$ indicates the initial value\footnote{The initial value of $\mu_{e_R}$ is chosen such that we can still use perturbation theory. As we will discuss shortly, the large value of $\mu_{e_R}$ is needed to amplify a small seed of hypermagnetic field amplitude. }. The conversion between $\mu_{e_R}$ and $n_{e_R}$ is simply done using $n_{e_R} = \mu_{e_R} T^2/6$. The BE governing the asymmetry of the right-handed electron is
\beq
\dot n_{e_R} +  3 H n_{e_R}  =  - \Gamma_{LR}  \left(n_{e_R} - n_{e_L} + \frac{n_\phi}{2}\right),
\label{eq:neR1}
\eeq 
where $n_i$ with $i = \{e_R, e_L, \phi,\cdots\}$ is the \textit{difference} between the number densities of a particle and its antiparticle\footnote{The ellipses mean that this definition applies to any particle.}, and $\phi$ represents the Higgs.Here, we use $H$ to represent a general Hubble rate (RD or EMD).  From the conservation of $B-L$ and neutrality of the plasma, we can find $n_{e_L}$ and $n_\phi$ in terms of $n_{e_R}$: 
\beq
n_{e_{L}} = - \frac{415}{962} n_{e_R}\ \ \ \ \ \ \  n_{\phi}  =  \frac{45}{481} n_{e_R}.
\eeq

To solve Eq.~\ref{eq:neR1} in the case of SC, it is more convenient\footnote{One can easily solve Eq.~\ref{eq:neR1} with respect to time: 
\begin{align*}
n_{e_R}^R(t) = &n^{0}_{e_R} \left(\frac{ t_\star}{t}\right)^{3/2} \nonumber\\
&\text{Exp}\left[- \frac{711 y_e^2 \sqrt{3 \Mpl}\left(\sqrt{t} - \sqrt{t_\star}\right)}{9620 (5 \pi)^{3/4} g_\star^{1/4}}\right].
\end{align*}} to do a change of variables:

\beq
\eta_{e_R}  = \eta_{e_R}^{0} \  \text{Exp}\left[ -\frac{2133 \ \Mpl \ y_e^2 \left(\frac{1}{T} - \frac{1}{\T}\right)}{19240\  \sqrt{5\  g_\star} \pi^{3/2}}\right],
\label{eq:etaRSC}
\eeq
 where $ \eta_{e_R} = n_{e_R}/s$, with $s$ being the entropy. Since we are only considering the temperatures greater than $\TEW$, $g_\star$ is a fixed number and it is equal to 106.75.  As Eq.~\ref{eq:etaRSC} shows, the comoving number density of the asymmetry is exponentially sensitive to the Yukawa coupling, and thus with a larger Yukawa coupling (e.g, muons), it would decrease much faster. 
Solving Eq.~\ref{eq:neR1} in the case of EMD is less trivial. Partially because of a more complicated Hubble rate, but also because the relation between temperature and time is a more complex relation: 
\beq 
T (t) \equiv \left(\frac{\rho_R (t)}{\frac{\pi^2}{30} g_\star}\right)^{1/4}. 
\eeq
However, one can find an approximate solution to Eq.~\ref{eq:neR1} in the case of EMD, by using the analytical equations given in Eq.~\ref{eq:analyticalrhos} and assuming $ \rho_\Phi \gg \rho_R$, when evaluating the Hubble rate (See Appendix ~\ref{app:anavsnum} for details). For the sake of accuracy, in the following, we work with the numerical solution of $n_{e_R}^M$. 
 
\begin{figure}
\centering 
\includegraphics[width=0.5\textwidth, height=0.43\textheight]{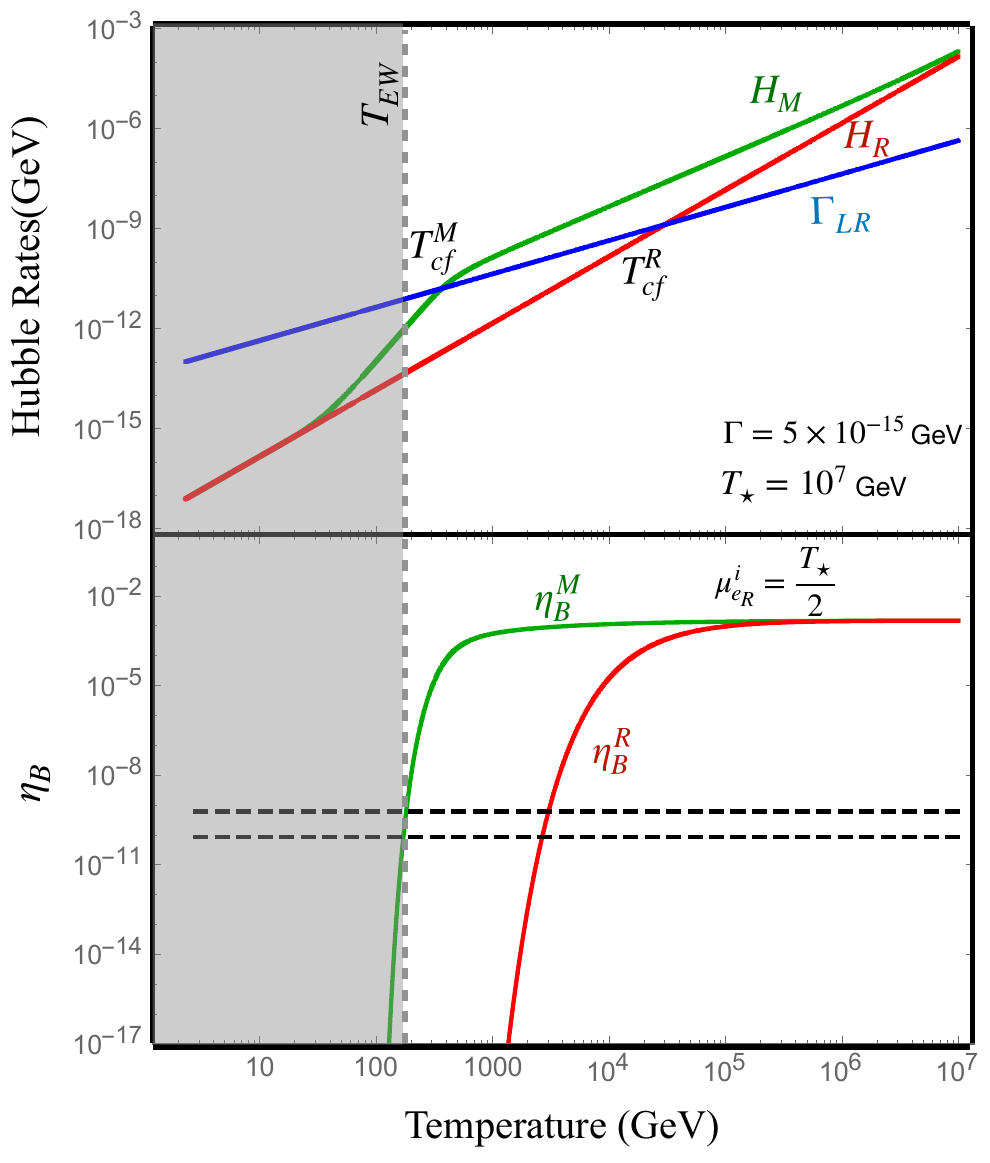}
\caption{In the upper panel, the Hubble rates $ H_\M$ (green) and $ H_\R$ (red) as well as $\Gamma_{LR}$ ( blue) are shown. In the lower panel, the evolution of $\eta_B$ according to Eq.~\ref{eq:neR1} is shown, where  the green line corresponds to $H= H_\M$ and the red line is when $H = H\R$. The initial asymmetry in the right-handed electron is assume to be  $\mu_{e_R}^0/ \T = 1/2$ to respect perturbation.  The horizontal dashed black lines is the approximate observed value of baryonic asymmetry, $\eta_B^{obs}$. The dashed gray line indicated $\TEW$, and the sphalerons become inactive for $T < \TEW$. Therefore, the value of $\eta_B$ freezes at $\TEW$. As the plots show, in SC, the chirality flip of the electron come into equilibrium earlier, and therefore the resulting asymmetry is much smaller than the observed value, $\eta_B^{R} (T = \TEW) \ll \eta_B^{obs}$. In the case of EMD, on the other hand, the asymmetry is kept for longer and $\eta_B^{M} (T = \TEW)$ matches the observed value. }
\label{fig:etaBprim}
\end{figure}

 The evolution of $\eta_B$ is shown\footnote{Even though Eq.~\ref{eq:neR1} describes $\eta_{e_R}$, we can find $\eta_B$ from the equilibrium conditions: $\eta_B =\frac{198}{481} \eta_{e_R}$.} in the lower panel of Fig.~\ref{fig:etaBprim}. As long as the Hubble rate is greater than the term involving $\Gamma_{LR}$, the comoving number density of asymmetry $\eta_B $ stays constant. However, once the rate of the chirality flip of the electron exceeds the Hubble rate, the asymmetry starts to decrease. As the upper panel of Fig.~\ref{fig:etaBprim} shows, $\Tx^{M} < \Tx^R$ meaning the asymmetry is preserved for longer. Another consequence of EMD is a dilution coming from the decay of $\Phi$ into the thermal bath. Therefore, we expect $\eta_B^M$ to decrease as a result of this dilution. This effect is shown by the rate at which $\eta_B$ decreases for $T < \Tx$. In the case of EMD, we see that the rate of the drop in $\eta_B$ is much greater than the SC case\footnote{There is a difference between $\Tx$ in the upper panel of Fig.~\ref{fig:etaBprim} and the temperature at which $\eta_B$ starts to decrease in the lower panel of Fig.~\ref{fig:etaBprim} coming from the different coefficients of $H$ and $\Gamma_{LR}$ in Eq.~\ref{eq:neR1}.}. 

\subsection{Primordial Asymmetry + Hypermagnetic field}
\label{sec:primeBY}
Here, we still assume that the fermions have a primordial asymmetry, but we include the effects of the hypermagnetic field as well. The baryon number violation in the SM is proportional to $\vec E_Y \cdot \vec B_Y$, where $\vec E_Y$ and $\vec B_Y$ are the hyper electric and hypermagnetic field, respectively. Therefore, it is crucial to study the effect of the hypermagnetic field on the evolution of the baryon asymmetry as well. ~\cite{Semikoz:2011, Dvornikov:2013, Kuzmin:1985mm, Zadeh:2018usa,Long:2014,Rubakov:1986am,Giovannini:1998,Giovannini:1998b,Joyce:1997, Khlebnikov:1988sr,Zadeh:2016nfk,Zadeh:2015oqf,Mottola:1990bz}. 

Interestingly, the observed large-scale magnetic field in the void between the galaxies is yet another puzzle in astrophysics and cosmology~\cite{Kronberg:1993vk,Kulsrud:2008,Harrison:1973zz}. The observation of magnetic field with similar magnitudes in a causally disconnect patches of universe, might be a hint of the existence of a primordial magnetic field~\cite{Semikoz:2011, Dvornikov:2013,Quashnock:1988vs, Kibble:1995,Sigl:1997,Vachaspati:1991nm,Enqvist:1993,Enqvist:1994,Olesen:1997,Baym:1996,Grasso:2001,Neronov:2010,Neronov:2009,Tavecchio:2011,Tavecchio:2010,Giovannini:1998b,Joyce:1997,Wolfe:2008}. If we insist that the existence of these magnetic fields roots in cosmology, a rough conservative estimate  suggest that we need to have a hypermagnetic field by the time of EWPT with an amplitude of $10^{19}~\G$ to explain the current magnetic field in the inter Galactic medium~\cite{Dvornikov:2013,Fujita:2016,Giovannini:1998b,Giovannini:1998,Giovannini:2013,Long:2016,Joyce:1997}.

The details of the inter-dependancies of the evolution of the hypermagnetic field amplitude (HMFA) and baryon asymmetry has been discussed in Ref.~\cite{Laine:2005,Appelquist:1981vg,Kajantie:1996,Joyce:1997,Zadeh:2015oqf,Zadeh:2018usa,Abbaslu:2019yiy,Elahi:2020pxl}. The Lagrangian describing $U(1)_Y$ in the Minkowski space is the following~\cite{Laine:2005,Joyce:1997,Zadeh:2015oqf,Kajantie:1996}.
\beq 
\Lc = - \frac{1}{4} Y_{\mu \nu} Y^{\mu \nu} - J_Y^\mu Y_\mu - c'_E \frac{\alpha'}{8 \pi} ( 2 Y\cdot B_Y),
\label{eq:MHlag}
\eeq
where $Y_{\mu \nu}$ is the field tensor of the hypercharge, and $\alpha'= 0.01$ is the structure constant of the hypercharge interaction. The first term in the Lagrangian~\ref{eq:MHlag} is the kinetic term,  $J_Y$ refers to the Ohmic current, and the term involving $c'_E$ is the Chern-Simons term. The value of $c'_E$ can be written in term of the chemical potential of the chiral fermions~\cite{Zadeh:2015oqf,Laine:2005}:
\begin{align}
c'_E &= \sum_{i=1}^{n_G} \left[-2 \mu_{R_i} + \mu_{L_i} - \frac{2}{3} \mu_{d_{R_i}}- \frac{8}{3} \mu_{u_{R_i}} + \frac{1}{3} \mu_{Q_i}\right]\nonumber\\
& = - 99/37 \mu_{e_{R}},
\label{eq:cep}
\end{align}
where the last equality comes from applying the equilibrium conditions. To study the evolution of the HMFA in the early universe, we must consider the expansion factor as well (the details are given in Refs. ~\cite{Abbaslu:2019yiy,Elahi:2020pxl}). The doing so, we can obtain the hyperelectric field:
 \beq
 \vec E_Y = \frac{1}{a \sigma} \vec \nabla \times \vec B_Y + \frac{\alpha'}{2 \pi \sigma} c'_E \vec B_Y - \vec v \times \vec B_Y,
 \label{eq:EY1}
\eeq 
where $a$ is the scale factor, $\sigma \simeq 100 T$ is the electrical hyperconductivity of the plasma, and $v$ is the the velocity of the plasma. Let us ignore the term containting the velocity of the plasma for now. This is justified, because the length scale of the change in the bulk velocity is much smaller than the correlation length of the hypermagnetic field. In other words, the infrared modes of the hypercharge are unaffected by the velocity of the plasma~\cite{Rubakov:1986am}. With this assumption, we can find the evolution of the hypermagnetic field as following:
 \beq
 \partial_t \vec B_Y + 2 H \vec B_Y = \frac{1}{a^2 \sigma} \nabla^2 \vec B_Y - \frac{\alpha'}{2 \pi a \sigma} c'_E  \vec \nabla \times \vec B_Y.
 \eeq
 One of the Maxwell equation of $U(1)_Y$ is that $\vec \nabla \cdot \vec B_Y =0$. Therefore, we can write $\vec B_Y = (1/a) \nabla \times \vec A_Y$, with $\vec A_Y$ being the vector potential. One particular non-trivial configuration of the vector potential that can result in fully helical hyeprmagnetic field\footnote{The magnetic field in the inter Galactic medium has been inferred to be helical, and therefore, a fully helical configuration is desirable\cite{Chen:2014qva}.} is 
 \cite{Lust:54,Dvornikov:2013,Chandrasekhar:56,Chandrasekhar:57,Zadeh:2018usa,Giovannini:1998,Giovannini:1998b,Zadeh:2016nfk,Zadeh:2015oqf}:
 \beq 
 \vec A_Y = \gamma (t) (\sin kz, \cos kz, 0),
 \label{eq:AY}
 \eeq
 where $\gamma(t)$ is the time-dependent amplitude of $\vec A_Y$, and $k$ is the comoving wave number. This configuration has been extensively studied in the literature~\cite{Lust:54,Dvornikov:2013,Chandrasekhar:56,Chandrasekhar:57,Dvornikov:2011ey,Dvornikov:2013,Zadeh:2015oqf,Abbaslu:2019yiy,Zadeh:2018usa,Zadeh:2016nfk,Giovannini:1999by}. It appears as if this configuration violates the assumption of homogeneity and isotropy condition. However, it can be shown that as long as the amplitude of the hypermagnetic field is smaller than $10^{23} \G$, the magnetic pressure ($B_Y^2/8\pi$) is orders of magnitude smaller than the fluid pressure. Thus, this violation of homogeneity and isotropy is negligible~\cite{Pavlovic:2016gac,Abbaslu:2020xfn,Elahi:2020pxl}.

With the configuration given in Eq.~\ref{eq:AY}, we can solve for hypermagnetic field. Notice that since the time dependent part of $\vec A_Y$ has been factored out of the vector part, the time dependent part of $\vec B_Y$ can similarly be separated, $B_Y(t) = k/a \gamma(t)$ (The details are given in Ref.~\cite{Elahi:2020pxl}). Therefore, we can find the BE of the amplitude of the hypermagnetic field
 \beq
\partial_t  B_Y + 2 H B_Y = -\frac{k'}{\sigma}  B_Y \left(k' + \frac{\alpha'}{2 \pi } c'_E  \right),
\label{eq:By}
 \eeq 
 where $k' = k/a \simeq kT$ . To continue the discussion, it is useful to note that  $-\frac{k^{'2}}{\sigma}  B_Y$ leads to a decrease in $B_Y$, where as $-\frac{\alpha' k'}{2 \pi \sigma} c'_E$ leads to an increase in $B_Y$.  As Eq.~\ref{eq:By} shows, if $\mu_{e_R}/T \simeq 1$, then $B_Y$ gets amplified. On the other hand, if $\mu_{e_R} =0$, it is \textit{not} generated by the hypermagnetic field. Similarly, the large HMFA has an effect on the evolution of baryon asymmetry as discussed in~\cite{Zadeh:2015oqf,Zadeh:2016nfk,Elahi:2020pxl}:
 \begin{align}
\dot n_{e_R} &+  3 H n_{e_R}  =  - \Gamma_{LR}  \left(n_{e_R} - n_{e_L} + \frac{n_\phi}{2}\right) \nonumber\\
&+ \frac{ \alpha' }{ \pi} \vec E_Y \cdot \vec B_Y, 
\label{eq:erBy}
\end{align}
 where 
 \beq
 \vec E_Y \cdot \vec B_Y  = \frac{ B^2_Y}{\sigma} \left( k' + \frac{\alpha'}{2 \pi }c'_E \right).
\label{eq:EB}
 \eeq
 
 Notice that if $B_Y = 0$, the asymmetry in the right-handed electron cannot generate a non-zero $B_Y$. However, as long as both $\mu_{e_R}$ and $B_Y$ are non-zero, they have a non-trivial influence on each other. The intricate connection between HMFA and baryon asymmetry has been discussed in Ref.~\cite{Zadeh:2015oqf,Zadeh:2016nfk}. As Eq.~\ref{eq:EB} shows, the existence of a non-zero $B_Y$ has two contributions on the evolution of $n_{e_R}$: one that is proportional to $k' B_Y^2$ which leads to an increase in $n_{e_R}$, and another one proportional to $ c'_E B_Y^2$ that leads to a decrease in the asymmetry. Therefore, a combination of large $B_Y$ and small $n_{e_R}$ is necessary for $B_Y$ to save $n_{e_R}$ from decreasing exponentially.
 
\begin{figure}
\centering 
\includegraphics[width=0.5\textwidth, height=0.43\textheight]{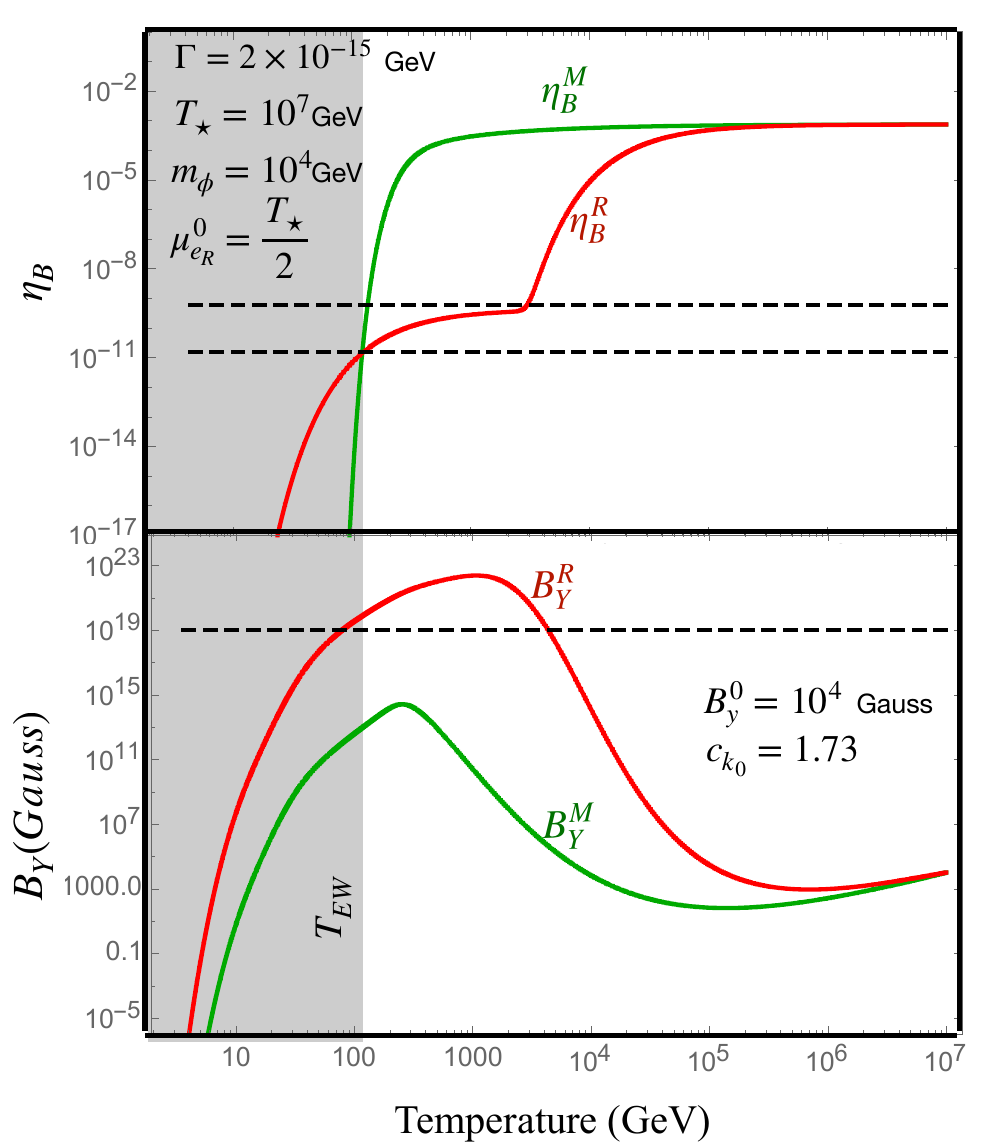}
\caption{In the upper panel, the evolution of $\eta_B$ and in the lower panel the evolution of $B_Y$ as a function of temperature for the benchmark indicated is presented. The large asymmetry in the right-handed electron causes $B_Y$ to increase. If $ \frac{\alpha' B^2_Y k'}{\pi \sigma}$ gets higher than  $ \Gamma_{LR}  n_{e_R}$, the asymmetry is prevented from diminishing. In EMD, due to the higher Hubble rate, $B_Y$ starts to increase at lower temperatures, and therefore it never gets large enough to compete with $\Gamma_{LR} n_{e_R}$.}
\label{fig:etaBprimBy}
\end{figure}

Fig.~\ref{fig:etaBprimBy} shows the temperature evolution of $\eta_B (T)$ and $B_Y(T)$ assuming 
 \begin{align}
 \Gamma = 2 &\times 10^{-15} \GeV, \hspace{0.1 in} T_\star = 10^7 \GeV, \hspace{0.1 in} \mu_{e_R}^0 = \frac{T_\star}{2}, \nonumber\\
 &B_Y^0 = 10^4 \G, \hspace{0.2 in} c_{k_0} \equiv \frac{k}{10^{-7}} =  1.73 ,
 \end{align}
in the two cases of SC and EMD.  Initially, $B_Y$ starts to decrease, because the Hubble rate is higher than the term with $c'_E$. As soon as $-\frac{\alpha' k'}{2 \pi \sigma} c'_E > 2 H$, then $B_Y(T)$ starts to increase. Since the Hubble rate in EMD is greater than the Hubble rate in SC, the decrease in $B_Y$ goes on for a longer time. The increase in $B_Y$ continues until $ k^{'2}/\sigma >  \left|\frac{\alpha' k'}{2 \pi \sigma} c'_E\right|$. As Fig.~\ref{fig:etaBprimBy} shows, if $B_Y$ increases to a large enough value, it can prevent $\eta_B$ from decreasing exponentially. The higher Hubble rate in EMD does not allow $B_Y$ to increase high enough, and thus $\eta_B$ in EMD continues to decrease exponentially\footnote{Furthermore, since the higher Hubble rate keeps $\Gamma_{LR}$ out of equilibrium for a longer time, $\mu_{e_R}$ in EMD is larger as $B_Y$ is increasing. This is another reason that $B_Y$ cannot help $n_{e_R}$ from decreasing exponentially.}.  As far as we are aware, this effect was not discussed in the literature so far. It is worth mentioning that any results we discuss here are highly sensitive to the initial conditions. As was illustrated in Appendix~\ref{app:anavsnum}, $\eta_B$ and $B_Y$ are exponentially sensitive to some of the free parameters.   Any small change in the initial conditions may change the final results drastically.
   
\subsection{Gradual increase of $\eta_{e_R}$ }
\label{sec:gradual}
In the following, we want to repeat the same analysis, but assume the injection of asymmetry into $e_R$ is gradual: from the decay of another particle, which we call $S$ for now. To keep our discussion general, we are going to be oblivious about the quantum numbers of $S$. The BE of $n_{e_R}$ due to the decay of $S\to e_R \cdots$ is
\begin{align}
\dot n_{e_R} &+  3 H n_{e_R}  =  - \Gamma_{LR}  \left(n_{e_R} - n_{e_L} + \frac{n_\phi}{2}\right) \nonumber\\
&+ \frac{ \alpha' }{ \pi} \vec E_Y \cdot \vec B_Y+ \Gamma_e n_S,
\label{eq:neRgrad}
\end{align}
where $\Gamma_e$ is the decay width of $S$ into $e_R$ and $n_S$ is the asymmetric number density of $S$. It is more convenient to express in terms of $\rho_S$: $n_S \equiv \eta_S \rho_S/ m_S$, where $\eta_S$ is a dimension-less parameter quantifying the amount of asymmetry in $S$. Therefore, the new parameters compared to the previous section are $$ \Gamma_e, \hspace{0.2 in} \eta_S, \hspace{0.2 in} m_S,  \hspace{0.2 in} \rho_S.$$
Even though $\rho_S$ is a free parameter, we will restrict it to to (1) $\rho_S = \rho_\R$, (2) $\rho_S = \rho_\Phi$, (3) $\rho_S = \rho_\R e^{- \frac{T- m_S}{T/2}}$. As mentioned earlier, we are not going to think about the explicit models where each of these densities are realized, though the model building for the second and third scenario are easily feasible(e.g, \cite{Chen:2019wnk}). 

\begin{figure}
\centering 
\includegraphics[width=0.5\textwidth, height=0.22\textheight]{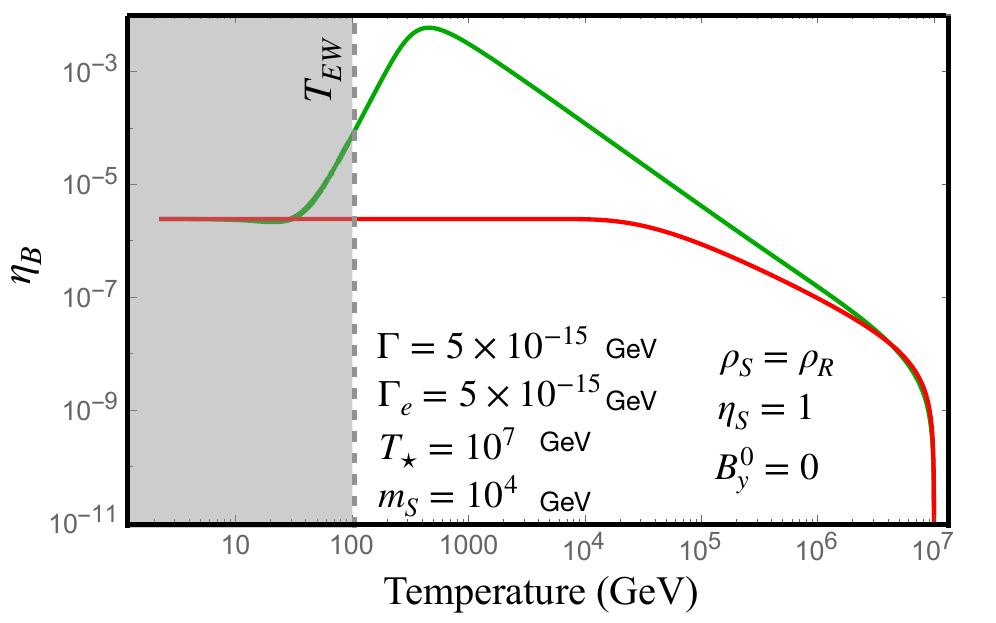}
\caption{The evolution of $\eta_B$ as a function of temperature for the two cases of SC (red) and EMD (green). In this plot, the asymmetry is due to the gradual decay of a particle $S$ into a right-handed electron for the benchmark shown. In SC, once the sphalerons become active, the rate of injection of asymmetry matches the wash-out rate, and thus $\eta_B$ becomes constant. In EMD, sphalerons become active much later, and thus $\eta_B$ increases for a long period. The decrease in $\eta_B$ is due to the dilution of the universe from the decay of $S$. Eventually, $\eta_B$ in the two cases become comparable.}
\label{fig:gradual1}
\end{figure}

The BEs of our interest (Eqs.~\ref{eq:Hubble},~\ref{eq:neRgrad}, and~\ref{eq:By}) are highly coupled and it is more convenient to show the effect of MD era through a numerical study of a few benchmarks. Let us start our analysis with a simple case where all of the parameters are the same in SC and EMD, except for the Hubble rate. Specifically, the first benchmark we consider is the following, 
\begin{align}
&\rho_S = \rho_\R,\hspace{0.3 in} \Gamma_e = 5\times 10^{-15} \GeV, \nonumber\\
& \eta_S = 1,\hspace{0.3 in}\ m_S= 10^4 \GeV.
\end{align}
In the case of EMD, we will assume $\Gamma_S = 5\times 10^{-15} \GeV$ as well. Fig.~\ref{fig:gradual1} shows the evolution of $\eta_B$ as a function of temperature for the specified benchmark. In the case of SC, $\eta_B$ becomes constant after sphalerons become active. That is because the rate of injection of asymmetry becomes comparable with the rate of wash-out of the asymmetry by the sphaleron. The larger Hubble rate delays the effects of sphalerons. Hence, $\eta_B$ increases for a longer period. Once $S$ decays and injects entropy to the universe, $\eta_B$ starts decreasing. Once sphalerons become active, $\eta_B$ decreases further until $\rho_S$ becomes negligible and it merges the SC case. This plot is assuming $B_Y^0 = 0$ and therefore the hypermagnetic field does not affect $\eta_B$.  

\begin{figure}
\centering 
\includegraphics[width=0.5\textwidth, height=0.43\textheight]{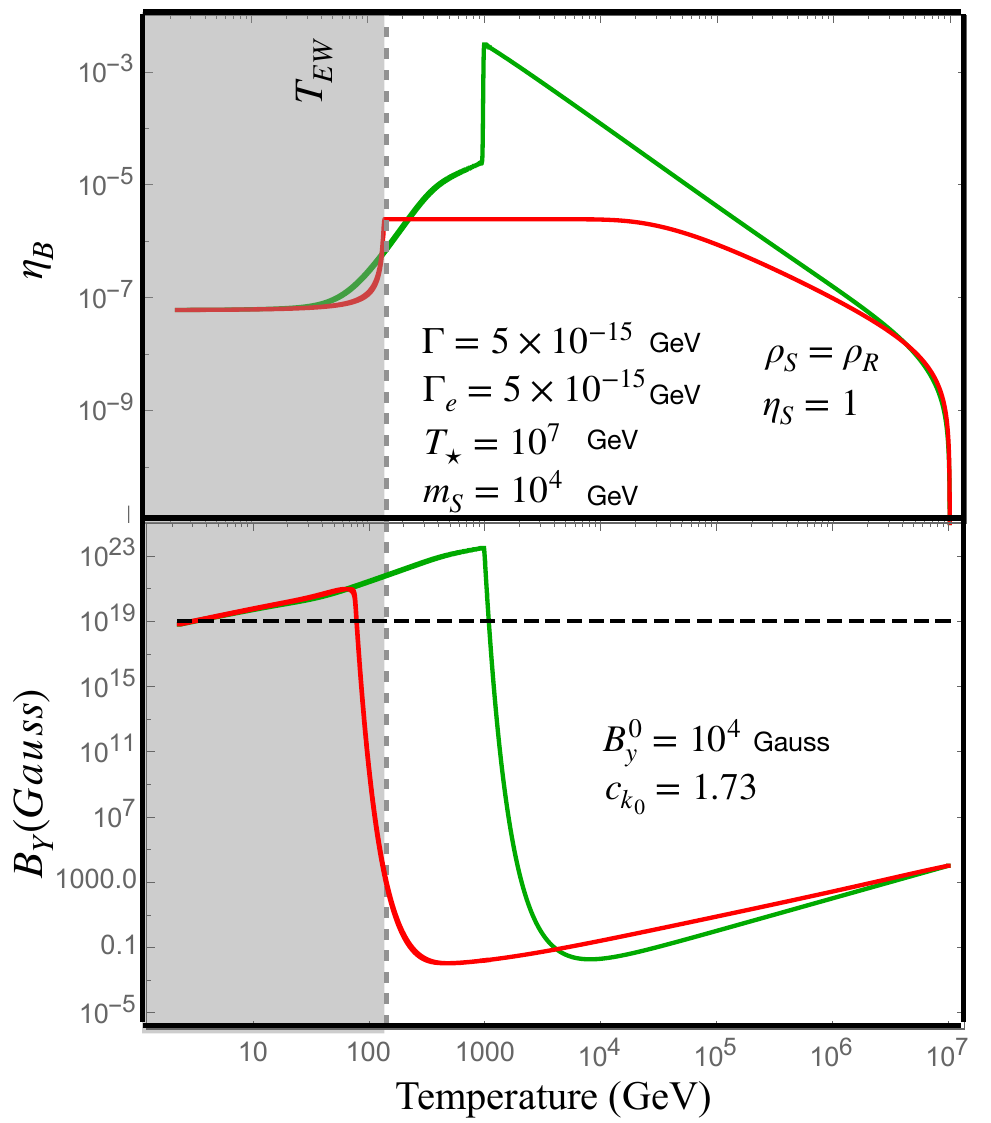}
\caption{The evolution of $\eta_B$ (upper panel) and $B_Y$ (lower panel) as a function of temperature for the two cases of SC (red) and EMD (green). In this plot, the asymmetry is due to the gradual decay of a particle $S$ into a right-handed electron for the benchmark shown. Here, we are highlighting the effect of $B_Y$ and its effect on the evolution of $\eta_B$. Since $\eta_B$ reaches a large value in EMD at higher temperatures, it leads to an earlier amplification of $B_Y$, such that by $\TEW$, we have reached the desired value of $B_Y$ in EMD, but $B_Y$ is very small in SC.}
\label{fig:gradual2}
\end{figure}

If we include the effect of $B_Y(T)$, the evolution of $\eta_B$ changes significantly. Recall that large $\eta_B$ leads\footnote{Large refers to larger than the Hubble rate.} to amplification in $B_Y$. A combination of large $B_Y$ and $\eta_B$, on the other hand, leads to a drop in $\eta_B$ (see Eq.~\ref{eq:EB}). The decrease in $\eta_B$ continues until either the Hubble rate dominates the process or $k' \simeq \frac{\alpha'}{2\pi} c'_E$. As we can see from Fig.~\ref{fig:gradual2}, since $\eta_B$ in EMD is larger than the SC case, the amplification of $B_Y(T)$ occurs faster, as opposed to Section~\ref{sec:primeBY}. The increase in $B_Y$ leads to a drop in $\eta_B$. For the case of EMD, $\eta_B$ decreases further due to the dilution of the universe. Interestingly, we see that when the asymmetry is injected gradually, the two cases of SC and EMD merge. That is if we were oblivious to EWPT. However, since after the EW phase transition, sphalerons become inactive, we are only sensitive to $\eta_B$ at $T_{EW}$.

\begin{figure}
\centering 
\includegraphics[width=0.5\textwidth, height=0.22\textheight]{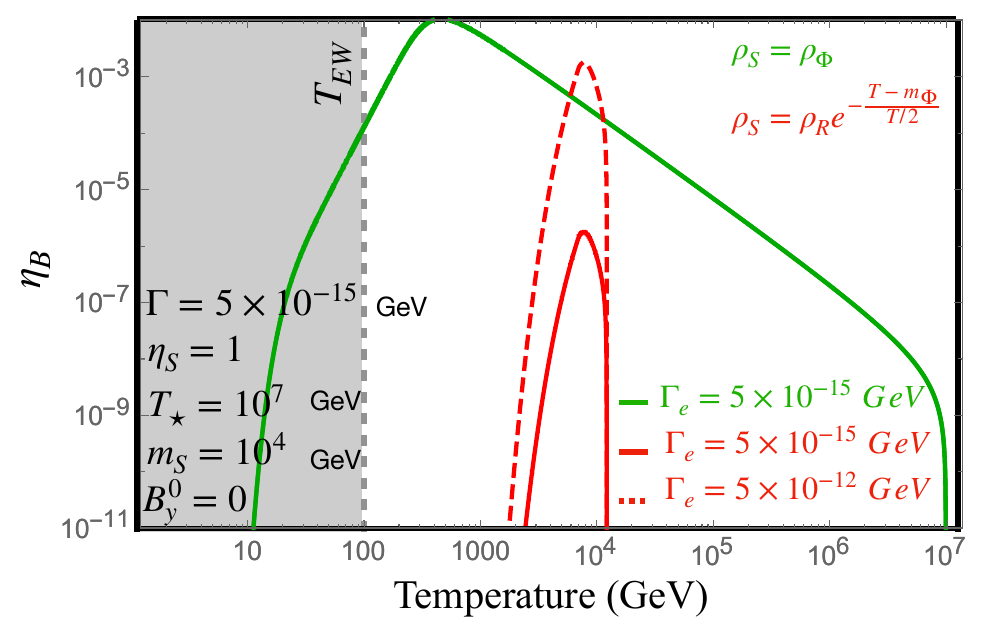}
\caption{The evolution of $\eta_B$ as a function of temperature for the specified benchmark is presented. The green line is assuming $\rho_S = \rho_\Phi$, and $\Gamma_e = \Gamma$. The solid red line corresponds to $\rho_S = \rho_\R  e^{- \frac{T- m_\Phi}{T/2}}$ and $\Gamma_e = \Gamma$. The decay width of $S$ in the case of SC is not limited from above. Therefore, the case where  $\rho_S = \rho_\R  e^{- \frac{T- m_\Phi}{T/2}}$ and $\Gamma_e = 10^3 \times \Gamma$ is also shown (dashed red).  }
\label{fig:gradual3}
\end{figure}

So far, we have only included the effect of the change in the Hubble rate. However, it is possible that the injection of asymmetry into $e_R$ is from the decay of $\Phi$. In other words, $ \rho_S = \rho_\Phi$. This case  has been discussed in the context of a specific model in Refs.~\cite{Chen:2019wnk} with $B_Y^0 = 0$ and ~\cite{Elahi:2020pxl}, where $B_Y^0 = 10^{-2} \G$. It is, however, difficult to compare such scenario with a similar scenario in SC. Maybe the closest reasonable assumption is to assume $\rho_S = \rho_\R e^{- \frac{T- m_\Phi}{T/2}}$ in SC and $\rho_S = \rho_\Phi$ in EMD. Thus, $\Gamma_e$ in EMD must be smaller than $\Gamma$; but there is no such requirement for the SC case and we can have $\Gamma_{e_{_{SC}}} \gg \Gamma$. To illustrate the effect of early MD in this case, let us consider the following benchmarks:
\begin{align}
&\eta_S = 1, \hspace{1.2 in} m_S = 10^4 \GeV,\nonumber\\ 
&\text{EMD}: \rho_S  = \rho_\Phi,\  \ \ \ \ \ \ \    \Gamma_e = 5\times 10^{-15} \GeV,  \nonumber\\
&\text{SC}:  \rho_S  = \rho_\R e^{- \frac{T- m_\Phi}{T/2}},\   \Gamma_e =\left\{\begin{array}{c} 5\times 10^{-15} \GeV\\ 5 \times 10^{-12} \GeV\end{array}\right..
\end{align} 
The evolution of $\eta_B$ as a function of temperature assuming $B_Y^0 = 0$, is shown in Fig.~\ref{fig:gradual3}. As the figure shows, $\eta_B$ increases due to the injection of asymmetry from $S$, and then it decreases because of the work of the sphaleron (and the dilution of the universe in the case of EMD). 

\begin{figure}
\centering 
\includegraphics[width=0.5\textwidth, height=0.43\textheight]{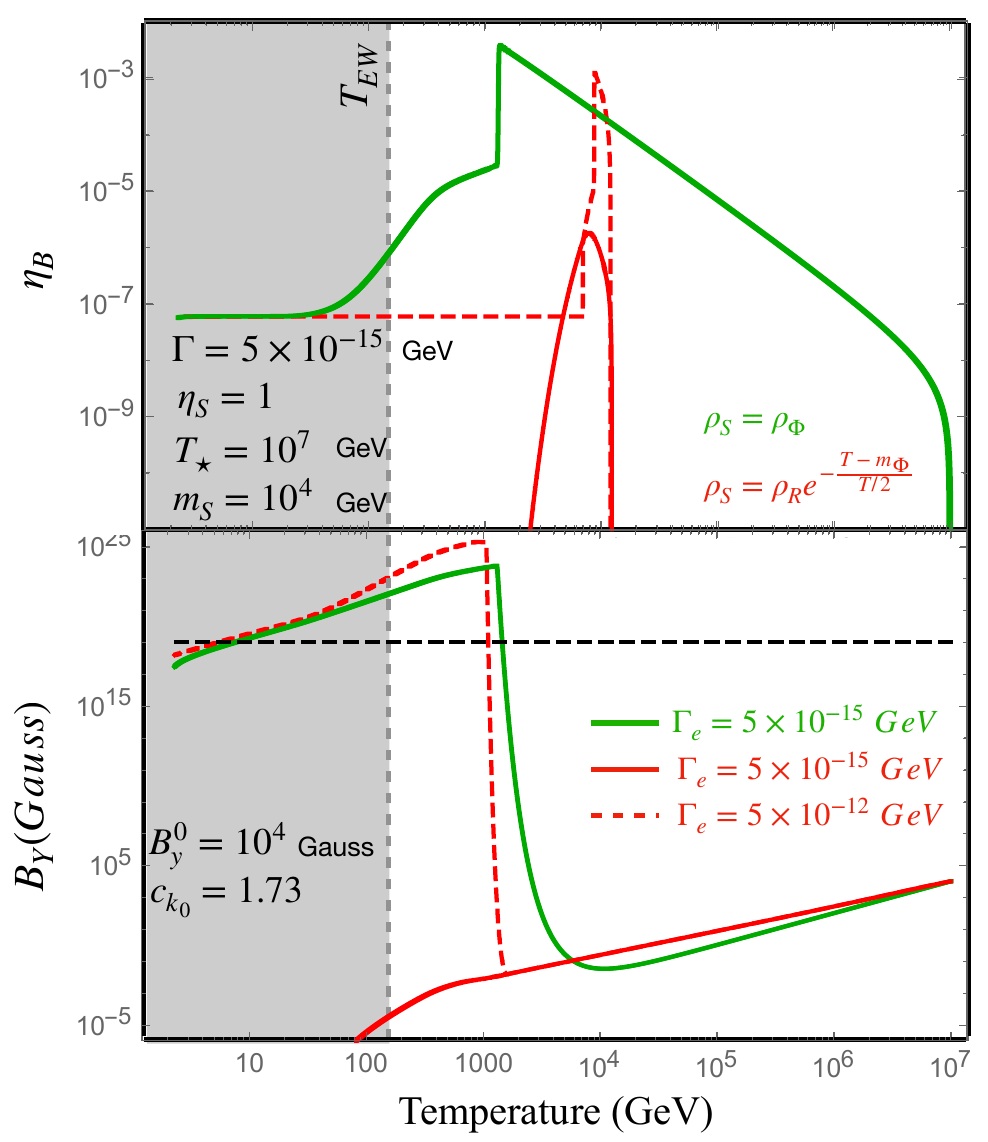}
\caption{The evolution of $\eta_B$ (upper panel) and $B_Y$ (lower panel) as a function of temperature for the specified benchmark is presented. The green line assumes $\rho_S = \rho_\Phi$ and the red lines correspond to $\rho_S = \rho_\R  e^{- \frac{T- m_\Phi}{T/2}}$. The $\Gamma_e$ in the solid lines is $\Gamma$ and the dashed line has $ \Gamma _e = 10^3 \Gamma$. As the plots show, if $\eta_B$ in the early universe hits a large value, it can cause an amplification in $B_Y$ which in return can save $\eta_B$ from diminishing.}
\label{fig:gradual4}
\end{figure}

A more interesting case is when we include the effect of $B_Y$ in the evolution of $\eta_B$ as well. As discussed earlier, large\footnote{ By large, we mean $\eta_B \gtrsim 10^{-4}$.} $\eta_B$ can lead to an amplification of $B_Y$ and in this process $\eta_B$ decreases to a smaller value. The amplified $B_Y$ on the other hand prevents $\eta_B$ from diminishing to zero. This behavior is shown in Fig.~\ref{fig:gradual4}. The case where $\eta_B$ does not hit a large value (solid red), end up having negligible $\eta_B$ \textit{and} $B_Y$. In the other cases, the value of $ \eta_B$ is such that it can lead to the amplification of $B_Y$. The values of $\eta_B$ and $B_Y$ at $\TEW$ in these cases are significant.

\section{Conclusion}
\label{sec:con}

The history of the early universe can have significant phenomenological impacts on the abundance of matter. In this paper, we studied the effect of early matter domination on the abundance of baryonic asymmetry ($\eta_B$). Since the evolution of $\eta_B$ is intimately connected with the amplitude of the hypermagnetic field, we studied the effect of the hypermagnetic field as well. 

Naively, one expects that the baryon asymmetry to be diluted in the case of early matter domination. Through considering a few benchmarks, we show that the effect of early matter domination can be highly non-trivial. In the case of early matter domination, $\eta_B$ can be orders of magnitudes greater, smaller, or even equal to $\eta_B$ in the standard cosmology. 

One key player in the evolution of $\eta_B$ is the weak sphaleron that leads to the washout of the asymmetry. Weak Sphalerons only act on left-handed fermions. An asymmetry in right-handed particles can be transferred to left-handed particles through Yukawa interaction. Hence, the right-handed electron because of its small Yukawa coupling has a special role in saving the asymmetry. If the right-handed electron gains an asymmetry, it will keep it for a relatively long time until its Yukawa interaction enters equilibrium. In standard cosmology, the Yukawa interaction of the right-handed electron is not slow enough to keep the asymmetry in the right-handed electron for a sufficiently long time. Studies have shown that assuming standard cosmology the sphalerons have enough time to wash out the asymmetry. In a universe with early matter domination, on the other hand, due to the change in Hubble rate, the sphalerons remain inactive for a longer period. Therefore, $\eta_B$ in a universe with a modified history can be many orders of magnitude greater than $\eta_B$ in standard cosmology. 

An asymmetry in right-handed electrons also affects the evolution of the hypermagnetic field. That is true only if the amplitude of the hypermagnetic field has a non-zero initial value. An asymmetry with values $\eta_B \gtrsim 10^{-4}$ may amplify a seed of hypermagnetic field amplitude. The grown amplitude will also feed to the asymmetry and save the asymmetry from diminishing. In a universe with an enhanced Hubble rate, the amplitude of the hypermagnetic field gets diluted. Whether the diluted magnetic field can get amplified or not is strongly dependent on the benchmark of the interest. Our numerical results show that in the scenario where the asymmetry in the right-handed electron is from a slow decay, the hypermagnetic field amplitude has a high chance of getting amplified. Hence, the evolution of both baryon asymmetry and the amplitude of hypermagnetic field are highly non-trivial and strongly depend on the benchmark of the interest. 

\section{Acknowledgement}
\label{sec:ack}
We would like to thank Gilly Elor, Enrico Morgante, Shiva Rostam Zadeh, Nicklas Ramberg, and Graham White for useful discussions. We are also grateful to IPM- Institute for Research in Fundamental Sciences for their support. The work of FE was supported by the Cluster of Excellence Precision Physics, Fundamental Interactions, and Structure of Matter (PRISMA+ EXC 2118/1) funded by the German Research Foundation(DFG) within the German Excellence Strategy (Project ID 39083149), and by grant 05H18UMCA1 of the German Federal Ministry for Education and Research (BMBF).  

\appendix
\section{Equilibrium Conditions:}
\label{app:EC}
The equilibrium values are derived from the following constraints (See \cite{Harvey:1990qw} and section 2 of  \cite{Cline:1994} for further details):
 \begin{itemize}
\item Due to flavor mixing the quark sector, all of the quarks regardless of their handedness have the same chemical potential
 \beq
\mu_Q \equiv \mu_{u_{R_i}} = \mu_{d_{R_i}} = \mu _{Q_i},
 \eeq
where $i =1,2,3$ indicates the generation of the quark. It is easy to see that 
\beq
\mu_B = \sum_i  2 \mu_{Q_i} +   \mu_{u_{R_i}} +  \mu_{d_{R_i}} = 12 \mu_Q. 
\eeq
It is worth mentioning that we are working in the limit of massless neutrinos, where there is no flavor mixing in the lepton sector. A more careful analysis where the lepton mixing is considered is beyond the scope of this paper. 
\item The work of weak sphalerons is so fast, that we can approximate it as being instantaneous:
\beq
9 \mu_Q + \mu_{L_1} + \mu_{L_2} + \mu_{L_3} = 0.
\eeq

\item The Yukawa interaction of all particles, except for $e_R$,  are in thermal equilibrium:
\begin{align*}
&\mu_{u_{R_i}} - \mu_{Q_i} = \mu_\phi, \ \ \ \ \ i = 1,2,3\\
& \mu_{d_{R_i}} - \mu_{Q_i} = -\mu_\phi,\ \ \ \ \  i = 1,2,3\\
& \mu_{e_{R_i}} - \mu_{L_i} =-\mu_\phi,\ \ \ \ \ i = 2, 3 \\
\end{align*}

\item. The plasma does not have any net hypermagnetic charge. 
\beq
Q = 6 \mu_Q -\mu_{R_1}- \mu_{L1} - 2 \mu_{L_2} - 2\mu_{L3} + 13 \mu_0 = 0
\eeq
\item $B-L$ is respected in each generation:
\beq
\frac{1}{3} \mu_B - L_i = C_i, 
\eeq
where $i$ indicates the generation again. Here, $C_i$s  are a constant and indicate the primordial values of $B/3 -L$ in each generation.  
\end{itemize}

\begin{figure}
\centering 
\includegraphics[width=0.43\textwidth, height=0.2\textheight]{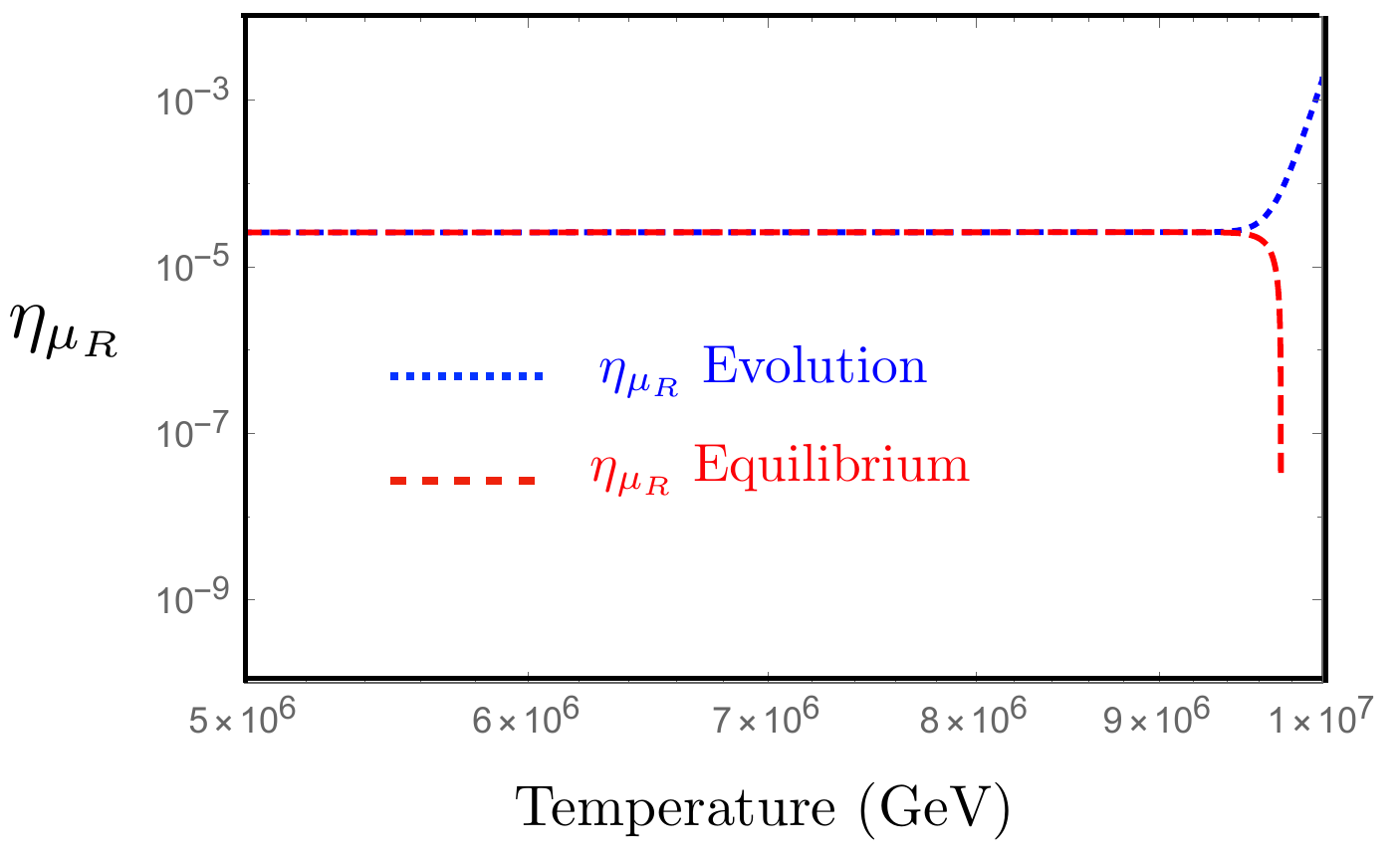}
\caption{The evolution of $\mu_R$ as a function of temperature. The initial value of $\mu_{\mu_R}^{0} $ is equal to $ \T/2$, and $\Gamma_\Phi = 5 \times 10^{-15} \GeV$. Notice that the right-handed muon approaches its equilibrium value before very rapidly.}
\label{fig:muR}
\end{figure}

These equilibrium conditions are such that we only have one unknown which is $\mu_{e_R}$. If we want to relax some of these conditions and end up with two unknowns, it is best to relax the condition where the Yukawa interaction of muon is in equilibrium. Similar to right-handed electron, right-handed muon is not involved in the work of strong and weak sphalerons, and it has a small yukawa couplings. In the following, we have relaxed this condition, and have looked at the evolution equation of the right-handed muon. We compare the evolution with the equilibrium value of the right handed muon in Fig.~\ref{fig:muR}. Notice that the right-handed muon quickly merges with its equilibrium value. In Fig.~\ref{fig:muR}, we had chosen the initial value of $\mu_{\mu_R}^{0} = \T/2$. For this figure, we have assumed EMD, with $\Gamma_\Phi = 5 \times 10^{-15} \GeV$. Notice that the right-handed muon quickly merges with its equilibrium value. 

\vspace{0.2 in}

\section{Analytic vs. Numeric:}
\label{app:anavsnum}
\begin{figure}
\centering 
\includegraphics[width=0.43\textwidth, height=0.2\textheight]{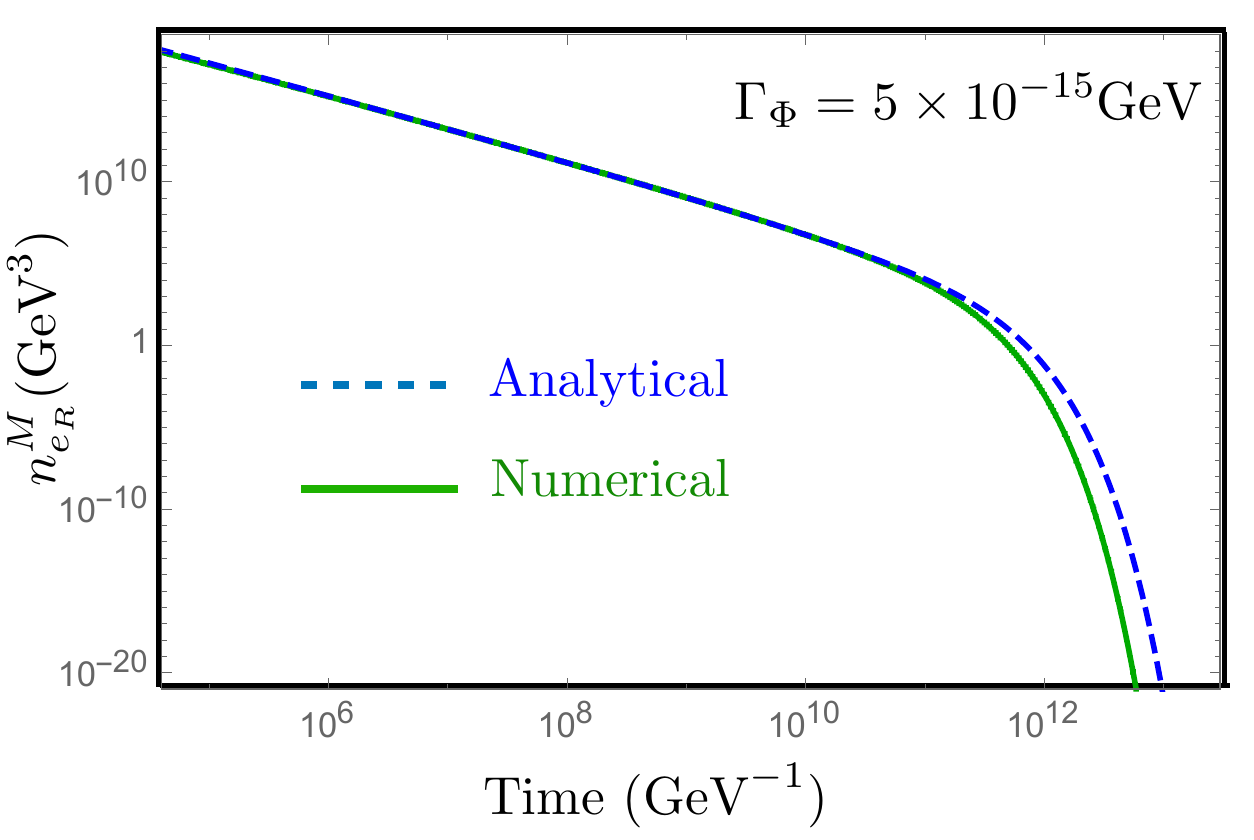}
\caption{The evolution of $n_{e_R}$ as a function of temperature, where the green line is the evolution obtained using numerical approach, and the blue dashed line is the analytical approach. Notice that the analytical approach is good estimate up to the time when the decay from $\Phi$ produces enough radiation that $\rho_\Phi \simeq \rho_R$, and thus violating our assumptions of the analytical approach. For this plot, we have assumed $\Gamma_\Phi = 5 \times 10^{-15} \GeV$.}
\label{fig:num}
\end{figure}

In this section, we compare the analytical solution to the numerical ones. 
Fig.~\ref{fig:num}  shows $n_{e_R}^M$ for the case where $\Gamma_\Phi = 5 \times 10^{-15} \GeV$, and the hypermagnetic field amplitude is turned off.The analytical solution (Eq.~\ref{eq:nMnum}) fits the numerical one up to the time when the decay of $\Phi$ becomes efficient.  

\begin{align}
& \text{\footnotesize {$n_{e_R}^M(t)  \simeq  n_{0} $}}\nonumber\\
& \text{\footnotesize {$\text{Exp} \left[ -2 \text{ExpIntEi} \left(-\frac{t \ \Gphi }{2}\right)+2  \text{ ExpIntEi} \left(-\frac{t_\star \Gphi }{2}\right)\right.$}} \nonumber\\
&\text{\footnotesize {$+\frac{79 \  y_e^2\  3^{1/4} } {48100 \  (10 \pi )^{3/4}}\left[ -8 \times 5^{3/4} \pi ( t \ T(t)  - t_\star \ \T ) \right.$}} \nonumber \\
&\text{\footnotesize {$ - \frac{ 25 \Mpl^2 t_\star^{2/3} \left( 5 + \frac{3 t^{5/3} \Gphi}{t_\star^2}\right)^{3/4} \text{ HG1F2} \left[\frac 1 5, \frac 3 4, \frac 6 5, - \frac{3 t^{5/3} \Gphi}{5 t_\star^{2/3}}\right]}{2 \pi t^{5/3} T(t)}$}}\nonumber\\
&\text{\footnotesize {$ \left. \left.- \frac{ 25 (t_\star \Mpl)^{1/2} \left( 5 + 3 t_\star  \Gphi \right)^{1/4} \text{ HG1F2}\left[\frac 1 5, \frac 3 4, \frac 6 5, - \frac{3 t_\star \Gphi}{5 }\right]}{(5 + 3 t_\star \Gamma)^{1/4}}\right] \right]$}},
\label{eq:nMnum}
\end{align}
 
 where ``ExpIntEi(z)" are the exponential Integral function $Ei(z) = - \int_{-z}^\infty \frac{e^t}{t} dt $, and ``HG1F2" are the hypergeometric function$_2 F_1$. As Eq.~\ref{eq:nMnum} illustrates, the baryon asymmetry is extremely sensitive to $\Gamma_\Phi$. Furthermore, since the baryon asymmetry is exponentially dependent on temperature,  once our assumptions breakdown, the analytical approximation may differ significantly from the actual $n_{e_R}$.

The analytical solution to $n_{e_R}$ and $B_Y$ is more complicated once we assume  a non-zero hypermagnetic field amplitude. Solving such highly coupled differential equations, even in the case of SC is non-trivial. We first change the differential to temperature, using\footnote{We emphasize that this relation is true for radiation domination, but it is only an approximation for EMD era.}: 
\begin{equation*}
\frac{d}{dt} = - H T \frac{d}{dT}.
\end{equation*} 
Then, we can get the following approximate solutions: 
\begin{align}
&\text{\footnotesize{$n_{e_R}^R =   \frac{{\bf n^{0}_{e_R}} T^3}{\T^3} \text{Exp}\left[ - \frac{21.33 \sqrt{5} y_e^2 \Mpl}{962 \pi^{3/2}  g_\star^{1/2}(T - \T)} \right]+ f(T) - f(\T)$}}\nonumber\\
& \text{\footnotesize{$ f(T) = \int_{T}^{\T} dT \ \frac{ 3\  {\bf c_{k_0}}\Mpl {\bf B_i}^2 T^2 }{5^7 \T^4} \times$}} \nonumber\\
& \text{\footnotesize{$ \text{Exp}\left[\frac{-3\  \Mpl {\bf c_{k_0}}^2}{40 \sqrt{5}\pi^{3/2}(g_\star)^{1/2} T} \right] \text{Exp}\left[\frac{3\sqrt{5} y_e^2 A  \Mpl }{2 \pi^{3/2}(g_\star)^{1/2} T} \right]$}} 
\end{align}
where $ A= 7.11/481$. 

\onecolumngrid
 Similarly, we can find the expression for $B_Y$ in SC: 
\begin{align}
& \text{\footnotesize{$B_Y^R  = B_Y^0 \text{Exp}\left[\frac{ \Mpl^3}{555000
\sqrt{5} A^6 \Mpl^8 \pi ^{3/2}( g_\star)^{3/2} T^5 \T^7 y_e^{12}} \right. 
\left(-{\bf c_{k_0}} \alpha' \left(-41625 A^6 {\bf c_{k_0}} \Mpl^6 T^3 \T^4 \left(-g_\star T \T^3 +g_\star T^2 \T^2\right)y_e^{12} \right) \right) $}}\nonumber\\
&\text{\footnotesize{$ +22 \sqrt{g_\star} \sqrt{\pi } \left(-\text{Exp}\left[\frac{3 \sqrt{5} A \Mpl (T-\T) y_e^2}{2 \sqrt{g_\star}
\pi ^{3/2} T \T}\right] +g_\star T^5 \T^3 \left(256 g_\star^{5/2} \pi ^{15/2} \T^4-384 \sqrt{5} A g_\star^2 \Mpl \pi ^6 \T^3 y_e^2\right.\right.
 +1440 A^2 g_\star^{3/2} \Mpl^2 \pi ^{9/2} \T^2 y_e^4$}}\nonumber\\
&\text{\footnotesize{$\left.  -720\sqrt{5} A^3 g_\star \Mpl^3 \pi ^3 \T y_e^6 +1350 A^4 \sqrt{g_\star} \Mpl^4 \pi ^{3/2} y_e^8 -2025 \sqrt{5} A^5 \Mpl^5 \eta_{e_R}^3 T^3 \T y_e^{10}\right)+\Mpl \left(-384 \sqrt{5} A g_\star^3 \pi ^6 T^4 \T^7 y_e^2 \right.
$}}\nonumber\\
&\text{\footnotesize{$-720 \sqrt{5} A^3 g_\star^2 \Mpl^2 \pi ^3 T^2 \T^7 y_e^6+1350 A^4 g_\star^{3/2} \Mpl^3 \pi ^{3/2} T \T^7 y_e^8+ 405 \sqrt{5} A^5 g_\star \Mpl^4 \T^7 y_e^{10}+405 \sqrt{5} A^5 g_\star \Mpl^4 T^5  \T^2  \left(1-5  \ {\bf n_{e_R}^0} \T^2\right) y_e^{10}
$}}\nonumber\\
&\text{\footnotesize{$
\left. \left. \left.+32 g_\star^{5/2}\Mpl \pi ^{9/2} T^3 \T^5 \left(8 \pi ^3 \frac{g_\star T^2 \T^2 }{\Mpl^2}+45
A^2 \T^2 y_e^4\right)\right)\right)\right)
\left.+555000 \sqrt{5} A^6 g_\star^{3/2} \Mpl^5 \pi ^{3/2} T^5 \T^7 y_e^{12} \text{Log}\left[\frac{T^2}{\T^2}\right]\right]$}}
\end{align}

\twocolumngrid

\begin{figure}
\centering 
\includegraphics[width=0.43\textwidth, height=0.2\textheight]{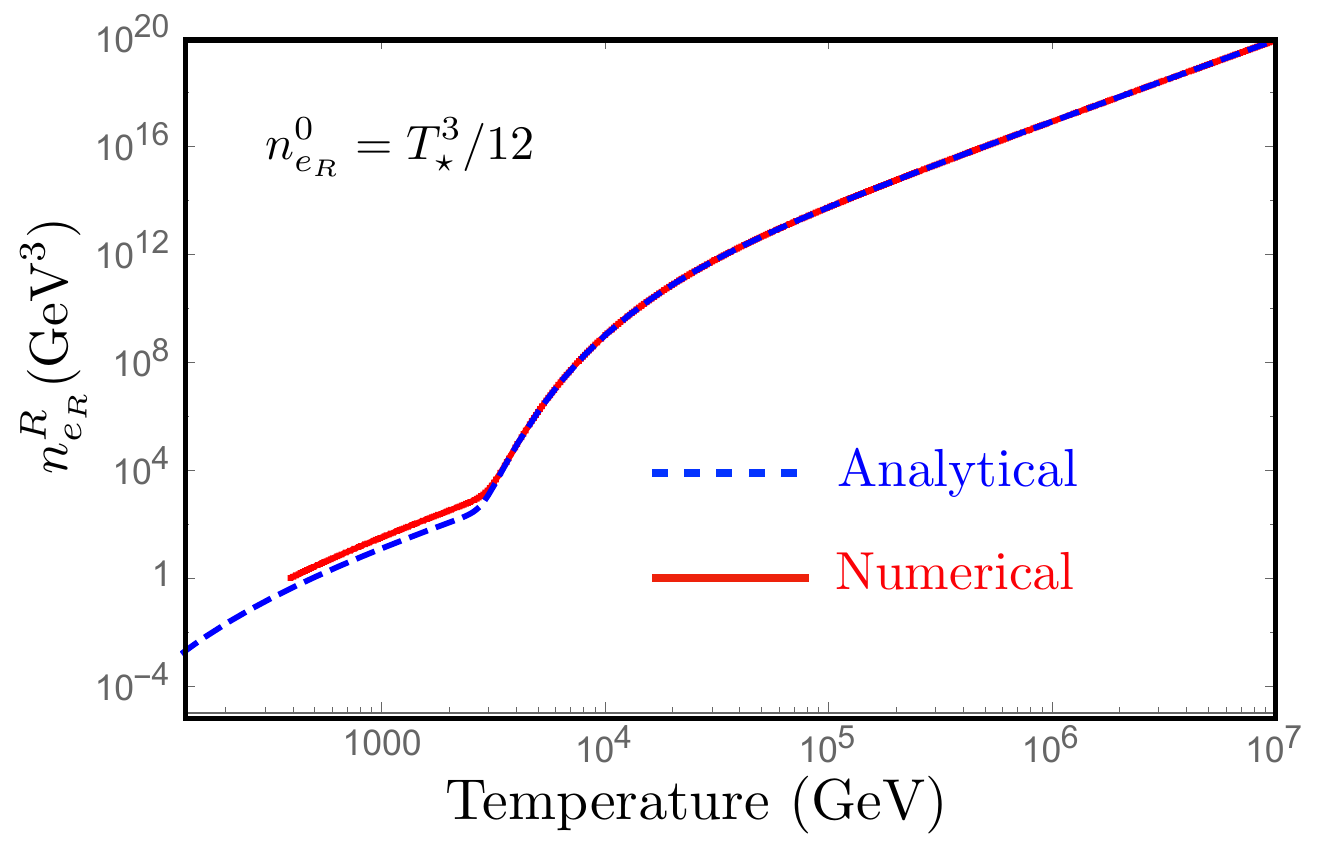} \hspace{0.2 in} \includegraphics[width=0.43\textwidth, height=0.2\textheight]{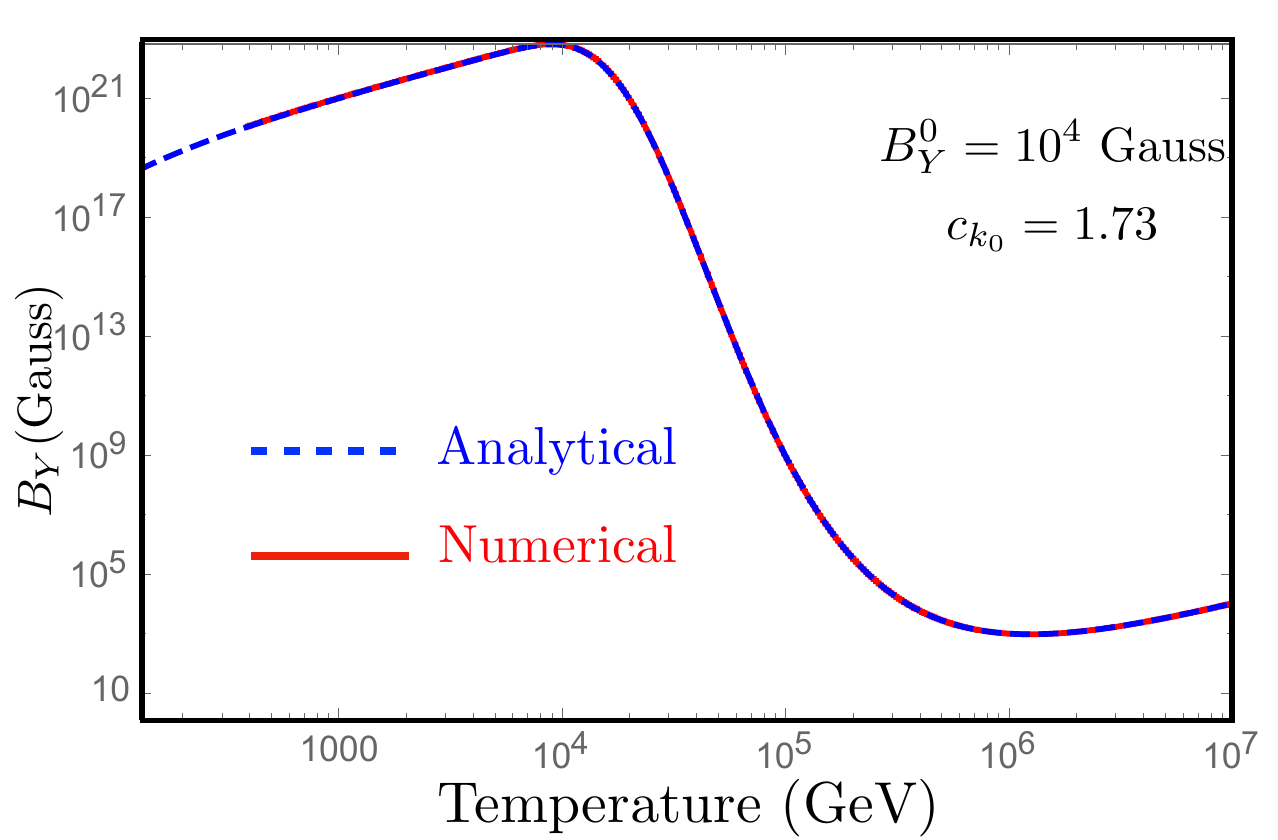}
\caption{The evolutions of $n_{e_R}$ and $B_Y$ as a function of temperature, where the red lines represent the numerical approach, and the blue dashed lines are the analytical solutions.}
\label{fig:nBYnum}
\end{figure}

As can be seen, the expressions get extremely complicated once we include the hypermagnetic field amplitude. The comparison between analytical solutions and numerical solution is shown in Fig.~\ref{fig:nBYnum}.  The analytical solution in EMD is much more complicated and could not be obtained.  Again, that is mainly because the relationship between temperature and time is more complicated. Nonetheless, it is easy to see that because of the larger Hubble rate during EMD era, the amplification of hypermagnetic field amplitude (specifically satisfying the condition $- \frac{\alpha' k'}{2 \pi \sigma} c_E' >  2 H$) occurs at a later time. Therefore, it is more difficult to amplify the hypermagnetic field amplitude to a desired value in the case of EMD. 

\onecolumngrid

\twocolumngrid
\bibliography{genesis}
\end{document}